\documentclass{iau} 
\usepackage{graphicx}
\graphicspath{{figures/}} 
\usepackage{natbib}
\title[MOCCA survey database and EGGC] 
{MOCCA survey database I: preliminary mock Extra Galactic
Globular Cluster observations}

\author[Agostino Leveque \& Miros\l aw Giersz]   
{Agostino Leveque$^{1}$ \and Miros\l aw Giersz$^{2}$}

\affiliation{Nicolaus Copernicus Astronomical Center, \\ Polish Academy of Sciences, Warsaw, Poland \\
$^1$ email:{\tt agostino@camk.edu.pl} \\[\affilskip] $^2$ email:{\tt mig@camk.edu.pl}
}

\pubyear{2019}
\volume{351}  
\jname{Star Clusters: From the Milky Way to the Early Universe}
\editors{A. Bragaglia, M.B. Davies, A. Sills \& E. Vesperini, eds.}
\begin{document}

\maketitle

\begin{abstract}
The photometric properties that we could observe for Extra-Galactic Globular Clusters (EGGCs) are the integrated light of the system and for nearby EGGCs it also is possible to measure both half-light radii and the color spatial distribution, e.g. for areas smaller and larger than the half-light radius. No information about the internal dynamical state of the system could be directly obtained from observations. On the other hand, simulations of Globular Clusters (GCs) can provide detailed information about the dynamical evolution of the system.

We present a preliminary study of EGGCs' photometric properties for different dynamical evolutionary stages. We apply this study to $12\, Gyr$ old GCs simulated as part of the MOCCA Survey Database. We determine the magnitudes in different bands from their projected snapshots using the Flexible Stellar Population Synthesis (FSPS) code and we measure the half-light radii from the surface brightness.
\end{abstract}

\firstsection 
\section{Introduction}
From observations, it is clear that globular clusters (GCs) are very common for all types of galaxies. They can provide a powerful diagnostic for galaxy formation, star formation in galaxies, galaxy interaction and mergers, and the distribution of dark  matter in galaxies. Indeed, galaxy-galaxy interactions can trigger major star-forming events, and the formation of massive star clusters. The properties of GCs systems in variuos galaxies can constrain the formation and the evolution of their host galaxies.

In the last decade of observation, the most interesting property of EGGC has been the discovery of color bimodality, 
 due principally to a metallicity difference, as it has been shown in recent spectroscopic studies \citep{Brodie2006}.

The $V-I$ color distribution for bright early-type galaxies show usually a blue peak at $V-I=0.95 \pm 0.02$, corresponding to $[Fe/H] \sim - 1.5$ and a red peak at  $V-I=1.18 \pm 0.04$, corresponding to $[Fe/H] \sim -0.5$ \citep{Larsen2001}. 
 Finally, the bimodality is not `universally observed', since for fainter galaxies only one single peak  is evident \citep{Peng2006}. Different scenarios have been suggested to explain the observed color distribution \citep{Ashman1992, Forbes1997, Cote1998}, although there is no consensus  concerning the origin of  such a distribution.

\begin{figure}[hbt!]
\begin{center}
\includegraphics[scale=0.13]{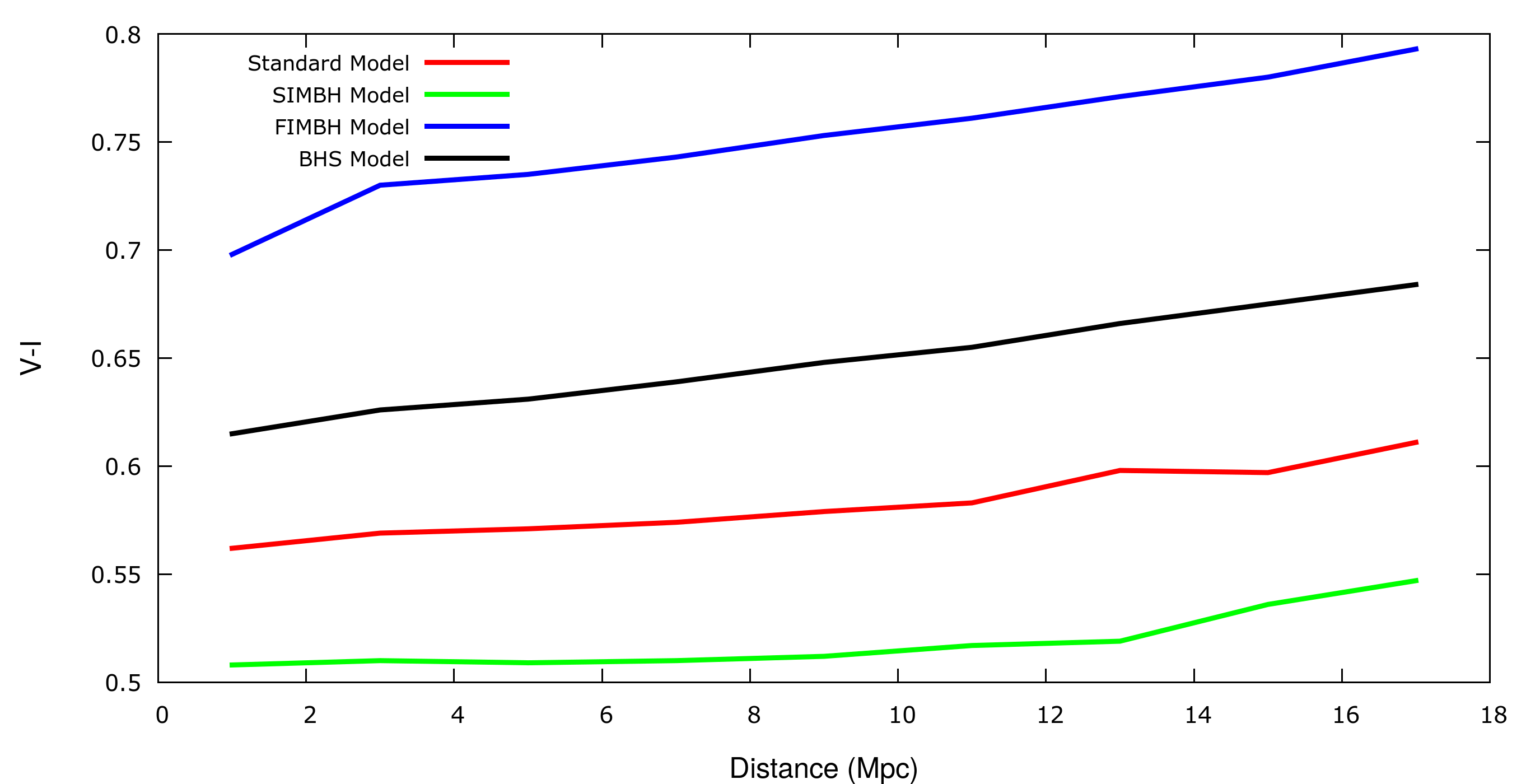}
\caption{The computed $V-I$ color as a function of the GC distance for chosen models.}
\label{B_V}
\end{center}
\end{figure}

\section{Models}
The ``MOCCA survey I database'' contains about 2000 models of GCs with different initial masses, structural and orbital parameters; it was shown that those models can be representative of the Milky Way GC population. More details are given in \cite{Askar2017}.

We selected models according to their dynamical state at $12\,Gyr$, dividing them into four groups:
\begin{itemize}
\item \textbf{Fast scenario} (FIMBH), presence of an IMBH (BH with masses $> 150 M_{\odot}$) formed before $1 \,Gyr$ \citep{Giersz2015};
\item \textbf{Slow scenario} (SIMBH), presence of an IMBH formed after $1\,Gyr$ \citep{Giersz2015};
\item \textbf{Black Hole Subsystem} (BHS), presence of a BH subsystem (number of BHs, $N_{BH} > 75$);
\item \textbf{Standard}, absence of an IMBH and of a BHS.
\end{itemize}

The total number of models is, respectively: \textbf{FIMBH:} 12; \textbf{SIMBH:} 8; \textbf{BHS:} 12; \textbf{Standard:} 15. All those models have metallicity $Z=0.001$. We do not impose any constrains on the initial conditions and the clusters' properities at $12\,Gyr$.

\section{Method}

\begin{figure}[hbt!]
\begin{center}
\includegraphics[scale=0.13]{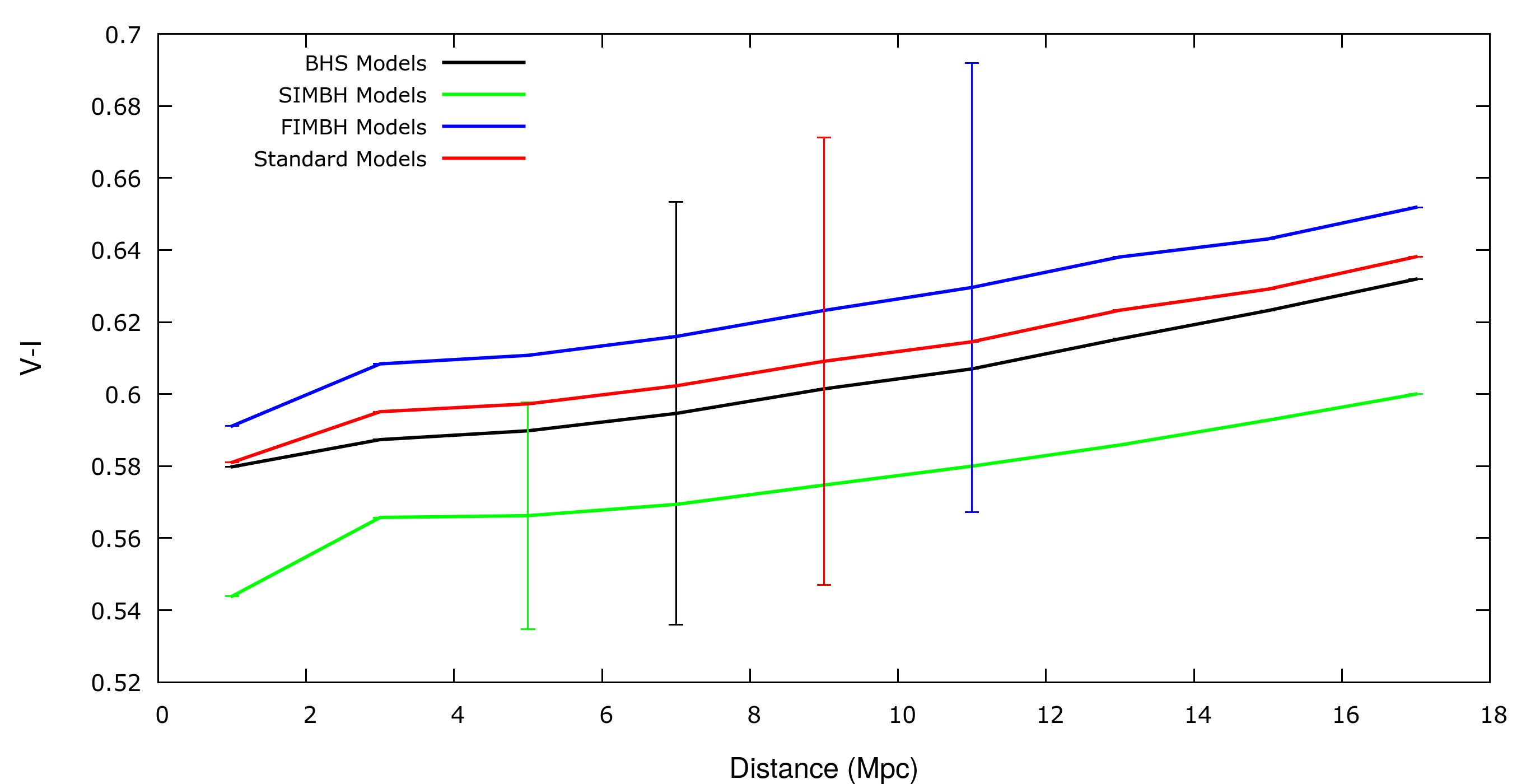}
\caption{The mean $V-I$ color over our sample as a function of the GC distance. The bar shows the standard deviation for each model.}
\label{Mean_B_V}
\end{center}
\end{figure}

For each selected simulation, we calculate the absolute magnitude of each star in the cluster using the FSPS code \citep{fsps1, fsps2}
 \footnote{FSPS is a stellar population synthesis code, returning the integrated spectra and the luminosity in different bands for a given stellar population, if its properties are given. We modify the code in order to obtain the star's photometry, if its radius, mass and luminosity are given, combining then all properties of GC's stars.}
 for five different optical bands (U, B, V, R, I). We put our model GCs at 9 distances (from $1 \, Mpc$ to $17\,Mpc$), taking in account the redshift effect for each distance.

Successively, we create a mock fits image using COCOA \citep{Askar2018}; the parameters are set to replicate the LBC cameras at the LBT observatory ($pixscale = 0.025''/pixel$, $seeing = 0.8''$, $gain = 2.0 \,photons/ADU$). We set no noise and no background in our mock images. The time exposure is set to $800\,s $ for V band, $1000\,s $ for B and U bands, $600\,s $ for R and I bands. 

\section{Results}
\subparagraph{\textbf{Results at $12\,Gyr$}}

In Fig. \ref{B_V} we show the obtained $V-I$ color for four of our models, one for each dynamical state group. We chose models with the largest spread in $V-I$ color. The range of $V-I$ color is of the order of $\sim 0.2$ dex, whereas there is not a strong distance dependency, with a color change of $\sim 0.5$ dex at the largest distance.

Fig. \ref{Mean_B_V}, instead, shows the mean value obtained from the $V-I$ colors of all chosen models.


From Figs. \ref{B_V} and \ref{Mean_B_V}, we can clearly see how the color differs depending on the cluster structure. Moreover, the models selected for this work have a wide range of initial condition parameters; the standard deviation around the average value obtained suggests that the initial GC parameters also have some influence on the color \citep{Askar2017}. We remark here that all of those models have the same metallicity; different GC histories could broaden the observed $V-I$ color distribution around the observed peaks.




\vspace{0.5cm}
\subparagraph{\textbf{Results at $9\,Gyr$}}

\begin{figure}[hbt!]
\begin{center}
\includegraphics[scale=0.13]{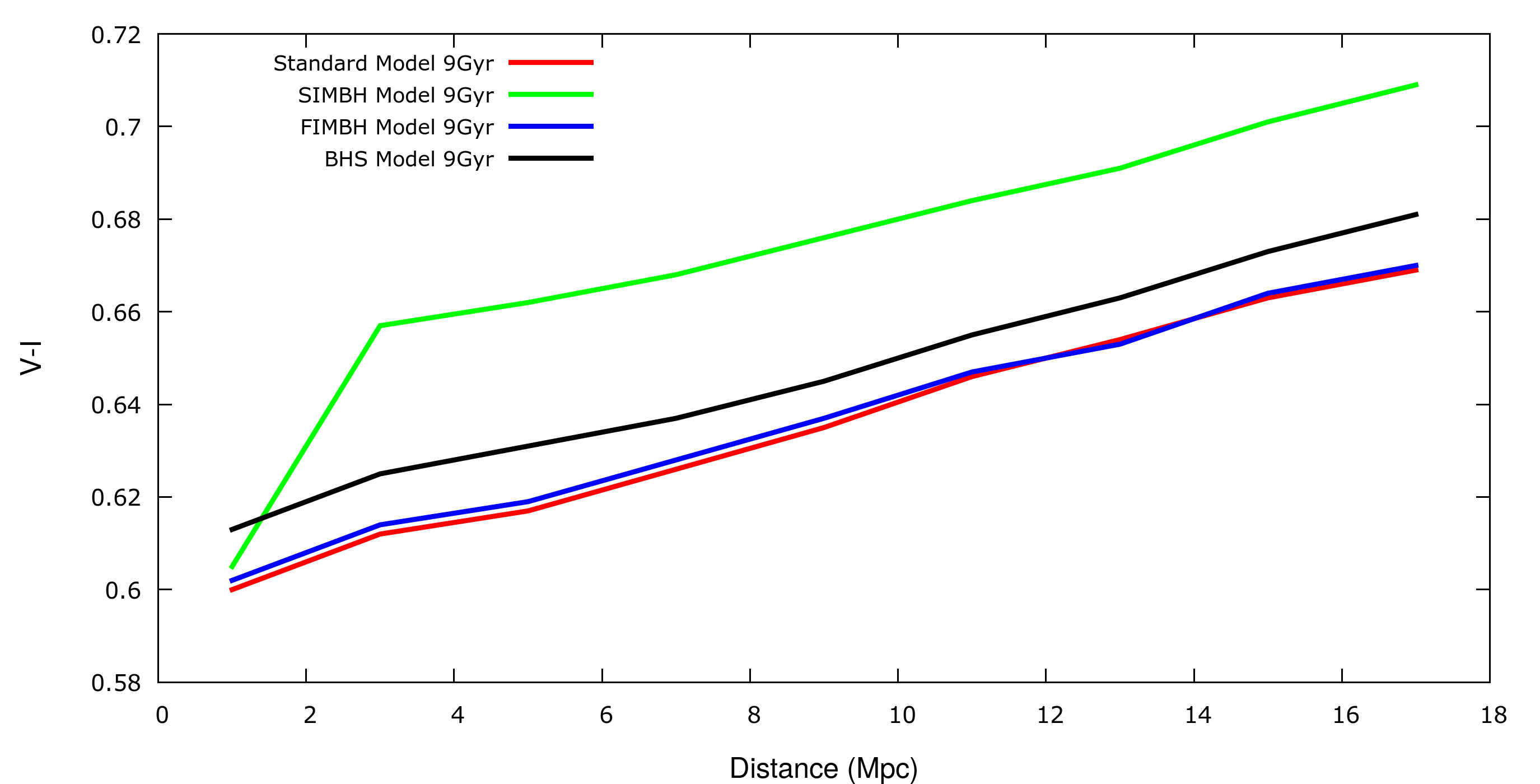}
\caption{The computed $V-I$ color at $9\, Gyr$ as a function of the GC distance for chosen models.}
\label{B_V9Gyr}

\end{center}
\end{figure}

In Fig. \ref{B_V9Gyr} we show the obtained $V-I$ color for four of our models at time $t=9\, Gyr$ of the GC lifetime. It can be seen how the color is also a strong function of the cluster age: the broadening is less pronounced for the models at $9\,Gyr$ (apart from the SIMBH model) with respect to the models at $12\,Gyr$. Furthermore, the typical value at $9\,Gyr$ results in being higher than the ones at $12\,Gyr$, due to stellar evolution. 
\vspace{0.5cm}

In conclusion, here we presented the very preliminary results of a simple analysis of the MOCCA models of GC evolution. It is just a proof of concept and in the future we will apply our method to a bigger sample of simulations.

\vspace*{2cm}
\section*{Acknowledgements}

MG and AL were partially supported by the Polish National Science Center (NCN) through the grant UMO-2016/23/B/ST9/02732.

\end{document}